\documentclass[a4paper,aps,prd,superscriptaddress,preprint]{revtex4}
\usepackage[utf8]{inputenc}
\usepackage{textcomp}
\usepackage{multirow}
\usepackage{amsmath}
\usepackage{amssymb}
\usepackage{graphicx}
\usepackage{hyperref}
\usepackage{orcidlink}

\makeatletter

\pdfpageheight\paperheight
\pdfpagewidth\paperwidth

\@ifundefined{textcolor}{}
{%
 \definecolor{BLACK}{gray}{0}
 \definecolor{WHITE}{gray}{1}
 \definecolor{RED}{rgb}{1,0,0}
 \definecolor{GREEN}{rgb}{0,1,0}
 \definecolor{BLUE}{rgb}{0,0,1}
 \definecolor{CYAN}{cmyk}{1,0,0,0}
 \definecolor{MAGENTA}{cmyk}{0,1,0,0}
 \definecolor{YELLOW}{cmyk}{0,0,1,0}
}

\usepackage{bm}
\usepackage{indentfirst}
\usepackage{float}
\usepackage{caption}
\usepackage{psfrag}
\usepackage{epstopdf}
\usepackage{diagbox}
\usepackage{orcidlink}

\@ifundefined{showcaptionsetup}{}{%
 \PassOptionsToPackage{caption=false}{subfig}}
\usepackage{subfig}
\makeatother

\begin{document}
\title{Extra-Dimensional de Broglie-Bohm Quantum Cosmology}

\author{F. A. P. Alves-J\'unior \orcidlink{0000-0002-9723-9717}}
\email{francisco.artur@univasf.edu.br}
\affiliation{Universidade Federal do Vale do S\~{a}o Francisco, \textit{Campus} Serra da Capivara, Piau\'i, Brazil}

\affiliation{Departamento de F\'isica, Universidade Federal de Campina Grande, Caixa
Postal 10071, 58429-900 Campina Grande, Para\'iba, Brazil}
\author{A. S. Lemos \orcidlink{0000-0002-3940-0779}}
\email{adiel@ufersa.edu.br}

\affiliation{Departamento de Ci\^{e}ncias Exatas e Tecnologia da Informa\c{c}\~{a}o, Universidade Federal Rural do Semi-\'Arido, 59515-000 Angicos, Rio Grande do Norte, Brazil}
\affiliation{Departamento de F\'isica, Universidade Federal de Campina Grande, Caixa
Postal 10071, 58429-900 Campina Grande, Para\'iba, Brazil}
\author{F. A. Brito \orcidlink{0000-0001-9465-6868}}
\email{fabrito@df.ufcg.edu.br}

\affiliation{Departamento de F\'isica, Universidade Federal de Campina Grande, Caixa
Postal 10071, 58429-900 Campina Grande, Para\'iba, Brazil}
\affiliation{Departamento de F\'isica, Universidade Federal da Para\'iba, Caixa Postal
5008, 58051-970 Jo\~{a}o Pessoa, Para\'iba, Brazil}

\begin{abstract}
In this work, we explore the de Broglie-Bohm quantum cosmology for a stiff matter, $p=\rho$,  anisotropic $n$-dimensional Universe. One begins by considering a Gaussian wave function for the Universe, which depends on the momenta parameters $q_1$ and $q_2$, in addition to the dispersion parameters $\sigma_1$ and $\sigma_2$. Our solutions show that the extra dimensions are stabilized through a dynamical compactification mechanism within the quantum cosmology framework. In this case, we find two distinct configurations for the dynamics of the extra dimensions. The first configuration features larger extra dimensions at the bounce, which subsequently undergo compactification to a smaller size. In contrast, the second configuration exhibits a smaller extra dimension at the bounce, evolving toward a larger, finite, and stabilized value. We also address the particular five-dimensional case where the Wheeler-DeWitt equation degenerates.

\end{abstract}

\maketitle

\section{Introduction}
Proposals suggesting a space-time with spatial extra dimensions date to the early 20th century \citep{Nordstrom:1914ejq}. The first extra-dimensional theory based on general relativity (GR) is the well-known Kaluza-Klein (KK) theory \citep{Kaluza:1921tu,Klein:1926tv}. This model presents a unified description of gravity and electromagnetism by postulating the existence of a five-dimensional space-time, in which the extra dimension should be compactified at a scale of the order of the Planck scale. Nowadays, extra-dimensional theories have attracted renewed attention owing to the braneworld models \citep{Randall:1999ee,Randall:1999vf,Antoniadis:1998ig,Arkani-Hamed:1998jmv,Arkani-Hamed:1998sfv}.
In this scenario, while Standard Model (SM) gauge interactions are trapped in the four-dimensional hypersurface (three-brane), gravity spreads out in the higher-dimensional space \citep{Randall:1999vf,Randall:1999ee}. Thus, these theoretical models achieve the unification between gravitational and gauge forces by postulating the existence of a supplementary space -- an essential ingredient of unification schemes like string theory \citep{Antoniadis:1998ig,Arkani-Hamed:1998jmv,Arkani-Hamed:1998sfv}.

Although research about extra-dimensional theories has received increasing attention in the braneworld framework, Kaluza-Klein models have also been exhaustively studied (see, \textit{e.g.}, \citep{Overduin:1997sri,wesson2006five} and references therein).
An exciting issue that appears in a cosmological model described by Kaluza-Klein-inspired extra dimensions is called dynamical compactification \cite{chodos1980has}. In this scenario, the small extra-dimensional length is a consequence of dynamical compactification, \textit{i.e.}, the cosmological evolution of the Universe imposes a tiny value to the extra dimension radius. Since then, many works have been published to address the compactification problem \cite{candelas1984calculation, van1986spontaneous,carroll2002classical,carroll2009dynamical}. Further, recent work examines how the dynamical compactification scenario of extra dimensions can explain the cosmic acceleration of Universe eras \cite{middleton2019anisotropic}.

It turns out that the quantum domain can offer additional insights into the cosmological implications of extra-dimensional theories. Therefore, it is known that Quantum Cosmology (QC) is a particular approach to Quantum Gravity (QG), which deals with the Wheeler-DeWitt equation aiming to explore the so-called wave function of the Universe.
Although the usual quantum mechanics is a successful theory of physics based on the well-established Copenhagen interpretation, such interpretation does not apply to QC since the Universe's wave function should not collapse. Furthermore, given the inconsistencies in the quantum Universe measurement process, alternative views must often be used in QC, such as the Many-Worlds (MW) interpretation proposed by Everett or even the de Broglie-Bohm (dBB) approach \cite{Novello:2009tn}. In the dBB interpretation, there is no wave function collapse, which is a key feature for QC. As a result, many works have followed this approach \cite{pinto2021broglie}. Ref. \cite{vicente2021quantum} shows that dBB QC applied to the Ho\v{r}ava–Lifshitz gravity can yield bounce and cyclic Universes. In turn, in Ref. \cite{vicente2023bouncing}, dBB formalism in the pre-inflationary era is investigated, while in \cite{maniccia2024quantum}, dBB corrections from the inflation spectrum are discussed.

In this context, the quantum cosmological scenario allows analyzing the Universe's fundamental properties addressing critical issues such as the stabilization of extra dimensions and dynamical compactification, for example \cite{vakili2006compactification}. Further, the Wheeler-DeWitt equation has been applied to the study of the Kaluza-Klein model in earlier papers such as \cite{rodrigo1984wheeler}. In its turn, Ref. \cite{darabi2000quantum} compares the Hartle-Hawking and Vilenkin tunneling approaches by considering a non-compact Kaluza-Klein Universe. Also, $n$-dimensional quantum cosmology has been studied to analyze the stabilization of extra dimensions as a signature change result \cite{jalalzadeh2003multi}. More recent results indicate that a Universe modeled by the metric $ds^2=-dt^2+a^2(dx^2+dy^2+dz^2)+b^2(\Sigma_{i=0}^n dw_i^2)$, where there are two scale factors -- $a(t)$ and $b(t)$ --, yields bouncing scenarios in the context of the quantum cosmology Everett interpretation \cite{alves2018quantum,pandey2019anisotropic},  where $\left<a\right>$ and $\left<b\right>$ depend on $n$. Therefore, this investigation sheds light on the issue of the dimensionality of space-time.

In this work, we aim to explore the de Broglie-Bohm interpretation for QC and show new scenarios, including intriguing flipped bounces for $b(t)$. Some consequences of our solutions are also discussed. Alternative to the compactification scenarios, extra-dimensional cosmologies could lead us to bounces \cite{das2017bouncing,anchordoqui2001brane}. In turn, the cosmological bounce models aim to avoid the singularity problems of the Universe, which could happen with or without an inflationary phase of the Universe. In the present paper, we address the question: ``What de Broglie-Bohm quantum cosmology can tell us about dynamical compactification?''. For this purpose, we model the Universe by a stiff matter. So, we solve the associated Wheeler-DeWitt equation and find its respective quantum Bohmian trajectories. We obtain several profiles of non-singular bouncing Universes in which the volume of the supplementary space is stabilized for a dynamical mechanism.

This article is structured as follows: In section $\textrm{II}$, we revisit the Wheeler-DeWitt equation for an $n$-dimensional anisotropic Universe. In section $\textrm{III}$, we find the Universe wave function and apply the dBB approach. We find two different classes of solutions, one of which is a family of solutions that considers an initial traveling wave function of the Universe, while the other admits an initial non-traveling wave function of the Universe. In this scenario, we discuss the bouncing profiles. Once the Wheeler-DeWitt equation derived is degenerated for $n=5$, we treat this case in section $\textrm{IV}$, where the same dBB-formalism. Finally, we present our final considerations in the last section.

\section{The Wheeler-DeWitt equation for anisotropic Universe}

In this section and in the following one, we revisit  the formulation for  an anisotropic Universe as proposed in \cite{alves2018quantum}. We start with the model described for the action 
\begin{equation}
 S=\int dx^{n} \sqrt{-g}\left(R+\omega g^{\mu\nu}\phi_{|\mu}\phi_{|\nu} \right),
\end{equation}
where $\varphi$ is a function of t, and $\omega$ is a positive constant.
This type of action appears in quantum cosmological models for Brans-Dicke gravitation theories \cite{fabris1999quantum,almeida2015quantum,almeida2021quantum}. Since $p=\rho$, this fulfillss the ultra-hard equation state, or stiff matter \cite{mukhanov2005physical}. We use the Chodos-like metric as
\begin{equation}
 ds^2=N(t)^2dt^2 -a(t)^2\sum^{3}_{i=0} dx_i^2
 -b(t)^2\sum^{n-4}_{i=0}dy_i^2.
\end{equation}
The direct calculation can lead us to get the Ricci scalar, 
$R^{(n)}=-\frac{6\ddot{a}}{aN^{2}}-\frac{6(n-4)\dot{a}\dot{b}}{abN^{2}}+\frac{6\dot{a}\dot{N}}{aN^{3}}-\frac{6\dot{a}^{2}}{a^{2}N^{2}}-\frac{2(n-4)\ddot{b}}{bN^{2}}+\frac{2(n-4)\dot{b}\dot{N}}{bN^{3}}-\frac{(n-4)(n-5)\dot{b}^{2}}{b^{2}N^{2}}$, and the associated  Hamiltonian according to \cite{alves2018quantum} is
\begin{equation}
 H=\frac{N}{(n-2)ab^{n-6}}\left[ \frac{(n-5)}{12}\frac{P^2_a}{b^2}+\frac{1}{(n-4)}\frac{P^2_b}{a^2}-
\frac{P_aP_b}{2ab}+\frac{(n-2)P^2_{\phi}}{4\omega a^2b^2}\right]. 
\end{equation}

For convenience, we proceed a canonical transformation
 where 
 $A=\ln a$ and $B=\ln b$. As a consequence, its corresponding momenta also change as $P_{A}=aP_{a}$ and $ P_{B}=bP_{b}$.
 In order to get Schr\"{o}dinger-like equation, we also change  $\phi$  following \cite{vakili2012scalar,farajollahi2010dynamics} as 
\begin{equation}
T=\frac{\phi}{P_{\phi}}, \hspace{1cm}
P_T=\frac{P^2_{\phi}}{2}.
\end{equation}
With this transformation, the $T$-parameter can measure the evolution of the system, with $P_T$ being its corresponding conjugate momentum. This approach allows us to deal with time in quantum cosmology in an alternative way instead of applying the Schutz formalism \cite{schutz1971hamiltonian}. Then, we finally get a suitable Hamiltonian
\begin{equation}
{H}=\frac{N}{2\omega a^3b^{n-4}}\left[ -\frac{\omega(n-5)}{6(n-2)}P^2_{A}-\frac{\omega}{n-2}P_{A}P_{B}
+\frac{\omega}{(n-2)(n-4)}P^2_{B}+P_{T}\right].
\end{equation}
 Choosing the conformal gauge  $N=2\omega a^3b^{n-4}$, we get $T=t$, and the time coordinate is recovered. Once we have the Hamiltonian, one may proceed with the canonical quantization ($P_A\rightarrow-i\frac{\partial}{\partial A}$, $ P_B\rightarrow-i\frac{\partial}{\partial B}$,
 and $P_T\rightarrow-i\frac{\partial}{\partial T}$). Then we the
  Wheeler-DeWitt for this anisotropic Universe can be written as
\begin{equation}\label{WDW}
 \left[-\frac{\omega(n-5)}{6(n-2)}\frac{\partial^2}{\partial A^2}+\frac{\omega}{(n-2)}\frac{\partial^2}{\partial A\partial B}
-\frac{\omega}{(n-2)(n-4)}\frac{\partial^2}{\partial B^2}\right]\Psi(A,B,T)
=i\frac{\partial}{\partial T}\Psi(A,B,T),
\end{equation}
That has a time-dependent Schr\"{o}dinger-like form. For instance, for a five-dimensional Universe, $n=5$, as pointed out, the equation degenerates and gets the form 
\begin{equation}
 \frac{\omega}{3}\left( \frac{\partial^2}{\partial A\partial B}-\frac{\partial^2}{\partial B^2}
\right)\Psi(A,B,T)=i\frac{\partial}{\partial T}\Psi.\label{psi_5}
\end{equation}

Throughout this article, we assume $\omega>0$, and here, we use the following transformation in coordinates 
\begin{equation}
x=\sqrt{\frac{6(n-2)}{|\omega|(n-5)}}A+\sqrt{\frac{(n-2)(n-4)}{|\omega|}}B, \label{u}
\end{equation}
\begin{equation}
 y=\sqrt{\frac{6(n-2)}{|\omega|(n-5)}}A-\sqrt{\frac{(n-2)(n-4)}{|\omega|}}B. \label{v}
\end{equation}
 In this new coordinate system, we arrive at a Schr\"odinger-like equation without the mixed derivatives.
\begin{equation}
\left[- \eta_{-}\frac{\partial^2 }{\partial x^2}-\eta_{+}\frac{\partial^2}{\partial y^2}\right]\Psi=i\frac{\partial \Psi}{\partial T},\label{schro}
\end{equation}
where $\eta_{\pm}=2\pm\sqrt{\frac{6(n-4)}{n-5}}$. This equation also corresponds to a free particle  with an effective mass, $m_{\eta_{-}}=\frac{1}{2\eta_{-}}$, in the $x$-direction, an effective mass $m_{\eta_{+}}=\frac{1}{2\eta_+}$, in the $y$-direction, which means that the anisotropy remains at the quantum level.

\section{De Broglie-Bohm Trajectories}
In 1951 David Bohm published two fundamental papers \cite{bohm1952suggested,PhysRev.85.180} with some criticism of the Copenhagen quantum mechanics view, and where it also described a new idea based on De Broglie pilot waves, the basis of his interpretation, which is radically different from the usual view since it is possible to calculate the particle trajectories. Nowadays, it is possible to find experimental pieces of evidence of the de Broglie-Bohm  trajectories
\cite{foo2022relativistic}.
 In quantum cosmology, this interpretation has been suitably applied (for a careful review, see \cite{pinto2013quantum}).
 
In summary, de Broglie-Bohm formulations start with imposing a given wave function profile as $\Psi=Re^{iS}$.  Applying this to an anisotropic Schr\"odinger equation \eqref{schro}, we can arrive at the following de Broglie-Bohm equations
\begin{equation}\label{eq:ODDB2}
    \frac{\partial R^2}{\partial t}+\nabla.\left(R^2\frac{\nabla S}{m}\right)=0,
\end{equation}
\begin{equation}\label{eq:ODDB1}
    \frac{\partial S}{\partial t}+\frac{(\nabla S)^2}{2m}+V-\frac{1}{2m}\frac{\nabla^2R}{R}=0.
\end{equation}
Whereas Eq. \eqref{eq:ODDB2} expresses the continuity equation, important quantities for this theory are obtained from the Hamilton-Jacobi-like equation \eqref{eq:ODDB1}, such as Bohmian trajectories, obtained for $\vec{p}=\nabla S$, and the quantum potential, $Q=\frac{1}{2m}\frac{\nabla^2R}{R}$. Similar quantum potential term can also appear in other contexts as a consequence of the wave function ansatz in the analog model \cite{anacleto2010acoustic}.
\subsection{Traveling Gaussian $\Psi_0$}
In this section, we aimed to solve the Schr\"odinger-like equation via propagation methods in order to get the Bohmian trajectories, in the same lines of \cite{delgado2020cosmological, pinto2009large}. For this purpose, we define $\bar{x}=x/\sqrt{\eta_{+}}$, and $\bar{y}=y/\sqrt{|\eta_{-}|}$. Then, we can rewrite the Eq. \eqref{schro} as follows:
\begin{equation}
    \left[ \frac{\partial^2 }{\partial \bar{x}^2}- \frac{\partial^2}{\partial \bar{y}^2}\right]\Psi=i\frac{\partial \Psi}{\partial T},\label{schro2}
\end{equation}
We note, however, that Eq. \eqref{schro2} is a Schr\"odinger equation that describes a free  particle with anisotropic mass, i.e., the mass in the $\bar{x}$-direction is $m_{\bar{x}}=-1/2$, whereas in the $\bar{y}$-direction, the mass is $m_{\bar{y}}=1/2$, that is a consequence of $\eta_{+}>0$ and $\eta_{-}<0$ for all $n$.

Now, we turn to the construction of $\Psi\left( \bar{x},\bar{y},T \right)$ from the initial state  $\Psi_0$. Then we start supposing that  the initial wave function of the Universe has  a Gaussian profile, given by
\begin{equation}
    \Psi_0(\bar{x},\bar{y})=\sqrt{\frac{2}{\pi\sigma_1\sigma_2}}
    \exp\left(-\frac{\bar{x}^2}{\sigma_1^2}+iq_1\bar{x}\right)\exp\left(-\frac{\bar{y}^2}{\sigma_2^2}+iq_2\bar{y}\right).
\end{equation}

This equation represents a traveling Gaussian function, where $q_1$ is the $\bar{x}$--momentum, while $q_2$ is the $\bar{y}$--momentum. In turn, $\sigma_1$ and $\sigma_2$ are the dispersions of the wave function (particle) in the respective directions. 
The appropriate propagator for this single particle that satisfies the Wheeler-DeWitt equation \eqref{schro2} has the form
\begin{equation}
K(\bar{x},\bar{y},\bar{x}',\bar{y}',T)= \frac{1}{4\pi T}\exp\Biggl[-i\frac{\left(\bar{x}-\bar{x}'\right)^2}{4T}+i\frac{\left(\bar{y}-\bar{y}'\right)^2}{4T} \Biggr].   
\end{equation}

Performing $\int^{+\infty}_{-\infty}\int^{+\infty}_{-\infty}K\Psi_{0} d\bar{x}' d\bar{y}'$, we arrive at the wave functions of the Universe for all times
\begin{equation}
    \Psi \left(\bar{x},\bar{y},T\right)=\frac{1}{p_1p_2}\frac{\sqrt{2\sigma_1\sigma_2}}{\sqrt{\pi}}\exp\left(\frac{-4(q_1+\frac{\bar{x}}{2T})^2T^2\sigma_1^{2}}{16T^2+\sigma_1^4}+\frac{-4(q_2-\frac{\bar{y}}{2T})^2T^2\sigma_2^{2}}{16T^2+\sigma_2^4}+iS\right),
\end{equation}
where 
\begin{align}
    &S=\frac{q_1^2T\sigma_{1}^4+q_1\bar{x}\sigma_1^4-4T\bar{x}^2}{16T^2+\sigma_1^4} +\frac{-q_2^2T\sigma_2^4+\bar{y}q_2\sigma_2^4+4T\bar{y}^2}{16T^2+\sigma_2^4},\\  
    &p_1=\sqrt{\left(i\sigma_1^2+4T\right)}, \qquad \text{and} \qquad p_2=\sqrt{\left(-i\sigma_2^2+4T\right)}.
\end{align}

We should emphasize that this particular Universe has a wave function that differs from that obtained from the Many-Worlds (MW) theory \cite{alves2018quantum}, where $\psi_{MW}=\sqrt{\frac{\sqrt{3(n-2)(n-4)}}{|\omega|\sqrt{\pi}}}\Big(\frac{\epsilon}{\epsilon^2+T^2}\Big)^{3/2}\bar{x}\bar{y}\exp\Big(-\frac{1}{4}\Big(
\frac{\bar{x}^2}{\epsilon+iT}+\frac{\bar{y}^2}{\epsilon-iT}\Big)\Big)$.
From the Broglie-Bohm pilot waveform, $\Psi=Re^{iS}$, one may determine the quantum potential, which would be ultimately responsible for yielding quantum effects in the early epochs of the Universe. In this context, it is expected that the classical limit of De Broglie-Bohm theory will be reached once the potential becomes negligible \cite{Novello:2009tn}. Thus, the quantum potential $Q$ is given by the expression:
\begin{equation}
    Q=\frac{4\left(q_1+\frac{x}{2T\sqrt{\eta_+}}\right)^2T^2\sigma_1^4}{(16T^2+\sigma_{1}^4)^2} -\frac{2\sigma_{1}^2}{16T^2+\sigma_1^4}-\frac{4\left(q_2-\frac{y}{2T\sqrt{|\eta_{-}|}}\right)^2T^2\sigma_2^4}{(16T^2+\sigma_{1}^4)^2} +\frac{2\sigma_{1}^2}{16T^2+\sigma_2^4}.
\end{equation}
As we can easily see, one has $Q\rightarrow 0$ for $T\rightarrow \pm\infty$, which ensures that quantum effects only occur at the origin $T\approx 0$. Furthermore, owing to the presence of the $Q$-potential, the classical trajectory will differ from the quantum one.

Now, rewritten $\bar{x}=x/\sqrt{\eta_{+}}$ and $\bar{y}=y/\sqrt{|\eta_{-}|}$, we can find the associated momenta by $p_x=\eta_{+}\left(\partial_x S\right)$ and $p_y=|\eta_{-}|\left(\partial_yS \right)$. These equations are equivalent to us writing $\dot{x}=\frac{1}{m_x}\frac{\partial S}{\partial x}$ and $\dot{y}=\frac{1}{m_y}\frac{\partial S}{\partial y}$, so that 
\begin{equation}
    \dot{x}=\frac{16Tx-2\sigma^4_1q_1\sqrt{\eta_+}}{16T^2+\sigma_1^4},
\end{equation}
\begin{equation}
    \dot{y}=\frac{16Ty+2\sigma_2^4q_2\sqrt{|\eta_{-}}|}{16T^2+\sigma_2^4},
\end{equation}
where $m_x=-1/2$ and $m_y=1/2$.

In what follows, we can now obtain the trajectory of the parameters $x$ and $y$:
\begin{equation}
    x=d_1\sqrt{\sigma_1^4+16T^2}-2q_1\sqrt{\eta_+}T, \label{x}
\end{equation}
\begin{equation}
    y=d_2\sqrt{\sigma_2^4+16T^2}+2q_2\sqrt{|\eta_{-}|}T, \label{y}
\end{equation}
where $d_{1}$ and $d_{2}$ are constants to be determined, related to the $x$-- and $y$--values in the bounce. Indeed, at $T=0$ one finds $d_{1}=x_{b}/\sigma_{1}^2$ and $d_{2}=y_{b}/\sigma_{2}^2$. Using expressions \eqref{u}, \eqref{v}, \eqref{x} and \eqref{y} we obtain the canonically transformed scalar factors
\begin{equation} A=\frac{1}{2}\sqrt{\frac{|\omega|(n-5)}{(n-2)6}}\left(d_1\sqrt{\sigma_1^4+16T^2}+d_2\sqrt{\sigma_2^4+16T^2}-2(\sqrt{\eta_+}q_1-\sqrt{|\eta_{-}|}q_2)T\right), \label{A}   
\end{equation}
\begin{equation}
     B=\frac{1}{2}\sqrt{\frac{|\omega|}{(n-2)(n-4)}}\left( d_1\sqrt{\sigma_1^4+16T^2}-d_2\sqrt{\sigma_2^4+16T^2}-2(\sqrt{\eta_+}q_1+\sqrt{|\eta_{-}|}q_2)T\right).  \label{B}
\end{equation}

Now, one can obtain the scale factors $a(T)=e^{A(T)}$, and $b(T)=e^{B(T)}$, so that:
\begin{align}
a(T)= & \exp\left[\frac{1}{2}\sqrt{\frac{|\omega|(n-5)}{(n-2)6}}\left(d_{1}\sqrt{\sigma_{1}^{4}+16T^{2}}+d_{2}\sqrt{\sigma_{2}^{4}+16T^{2}}\right.\right.\notag\\
 & \Biggl.\biggl.-2\left(\sqrt{\eta_{+}}q_{1}-\sqrt{|\eta_{-}|}q_{2}\right)T\biggr)\Biggr],
\end{align}
\begin{align}
b(T)= & \exp\left[\frac{1}{2}\sqrt{\frac{|\omega|}{(n-2)(n-4)}}\left( d_1\sqrt{\sigma_1^4+16T^2}-d_2\sqrt{\sigma_2^4+16T^2}\right.\right.\notag\\
 & \Biggl.\biggl.-2\left(\sqrt{\eta_{+}}q_{1}+\sqrt{|\eta_{-}|}q_{2}\right)T\biggr)\Biggr].
\end{align}

At this point, we can determine the Hubble parameters by defining, for ordinary (4D) coordinates, $H_a(T)=\dot{a}\left(T\right)/a\left(T\right)$, while for the extra dimensions, one has $H_b(T)=\dot{b}\left(T\right)/b\left(T\right)$, as follows:
\begin{align}
H_{a}(T)=\sqrt{\frac{|\omega|(n-5)}{(n-2)6}}\left(\sqrt{|\eta_{-}|}q_{2}-\sqrt{\eta_{+}}q_{1}+\frac{8d_{1}T}{\sqrt{\sigma_{1}^{4}+16T^{2}}}+\frac{8d_{2}T}{\sqrt{\sigma_{2}^{4}+16T^{2}}}\right), \label{Ha}
\end{align}
\begin{align}
H_{b}(T)=-\sqrt{\frac{|\omega|}{\left(n-2\right)\left(n-4\right)}}\left(\sqrt{\eta_{+}}q_{1}+\sqrt{|\eta_{-}|}q_{2}-\frac{8d_{1}T}{\sqrt{\sigma_{1}^{4}+16T^{2}}}+\frac{8d_{2}T}{\sqrt{\sigma_{2}^{4}+16T^{2}}}\right), \label{Hb}
\end{align}
where the dot denotes differentiation with respect to the time T. On its turn, by defining $H_{a}\left(+\infty\right)=\lim_{T\rightarrow+\infty}H_{a}\left(T\right)$ as well as $H_b(\pm \infty)= \lim_{T \to \pm \infty} H_b(T)$, we find the constants
\begin{equation}
    d_1=\frac{q_1\sqrt{\eta_{+}}}{2}+\frac{1}{4}\left[H_a(+\infty)\sqrt{\frac{(n-2)6}{|\omega|(n-5)}}+H_b(+\infty)\sqrt{\frac{(n-2)(n-4)}{|\omega|}}\right], \label{d1}
\end{equation}
\begin{equation}
    d_2=-\frac{q_2\sqrt{|\eta_{-}|}}{2}+\frac{1}{4}\left[H_a(+\infty)\sqrt{\frac{(n-2)6}{|\omega|(n-5)}}-H_b(+\infty)\sqrt{\frac{(n-2)(n-4)}{|\omega|}}\right]. \label{d2}
\end{equation}

In what follows, we must assume a condition of physical interest that leads to the compactification of the extra dimensions, represented by $H_b(+\infty)=0$. This condition will express the dynamic compactification of the extra dimensions. Furthermore, from the compactification condition and considering $T=0$ in Eq. \eqref{Hb}, one gets a relationship between $q_1$, $q_2$, $d_1$, and $d_2$, which may be expressed by the equation $d_1-\sqrt{\eta_+}q_1/2=d_2+\sqrt{|\eta_{-}|}q_2/2$. For the sake of simplicity, we can take $d_{1}=d_{2}$, and thus we get
\begin{equation}
q_{2}=-q_{1}\sqrt{\frac{\eta_{+}}{|\eta_{-}|}} \label{q2}.
\end{equation}

It is worth noting that Eq. \eqref{q2}, along with the condition $H_b(+\infty)=0$, ensures that the $b$-scale factor does not vanish or diverge for $T\rightarrow+\infty$, and so assumes a finite value. As a matter of fact, if $d_{1}>d_{2}$, $b(T)\overset{T\rightarrow\infty}{\rightarrow}\infty$, whereas for $d_{1}<d_{2}$, one gets $b(T)\overset{T\rightarrow\infty}{\rightarrow}0$. These cases have little or no physical interest. As we will see later, the ordinary Hubble parameter, $H_a\left(+\infty\right)$, contributes to the extra-dimensional scale factor.

Hence, the scale factors take the form
\begin{align}\label{a-final-n-dim}
a(T)= & \exp\left[\frac{q_{1}}{4}\sqrt{\frac{\eta_{+}|\omega|(n-5)}{(n-2)6}}\left(\sqrt{\sigma_{1}^{4}+16T^{2}}+\sqrt{\sigma_{2}^{4}+16T^{2}}-8T\right)\right.\notag\\
 & \Biggl.\biggl.+\frac{H_a(+\infty)}{8}\left(\sqrt{\sigma_{1}^{4}+16T^{2}}+\sqrt{\sigma_{2}^{4}+16T^{2}}\right)\Biggr],
\end{align}
\begin{align}\label{b-final-n-dim}
b(T)= & \exp\left[\frac{1}{8\sqrt{(n-2)(n-4)}}\left(\sqrt{\sigma_{1}^{4}+16T^{2}}-\sqrt{\sigma_{2}^{4}+16T^{2}}\right)\right.\notag\\
 & \Biggl.\biggl.\times\left(2\sqrt{\eta_{+}|\omega|}q_{1}+\sqrt{\frac{(n-2)6}{(n-5)}}H_a(+\infty)\right)\Biggr].
\end{align}
\begin{figure}[tb]
\subfloat[\label{fig1a}]{\includegraphics[scale=0.8]{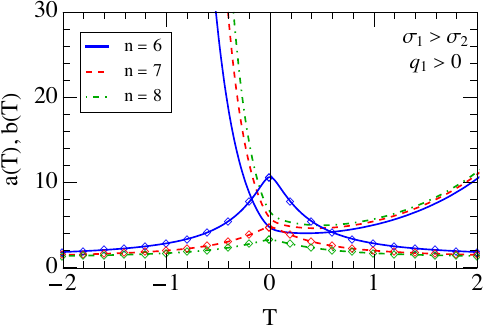}} \hspace{0.2cm} \subfloat[\label{fig1b}]{\includegraphics[scale=0.8]{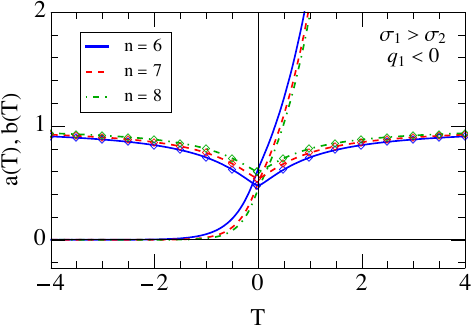}}

\subfloat[\label{fig1c}]{\includegraphics[scale=0.75]{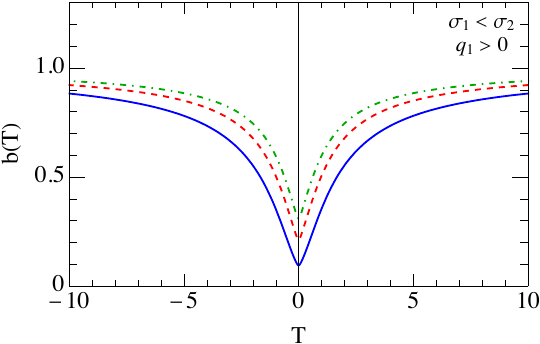}} \hspace{0.2cm} \subfloat[\label{fig1d}]{\includegraphics[scale=0.75]{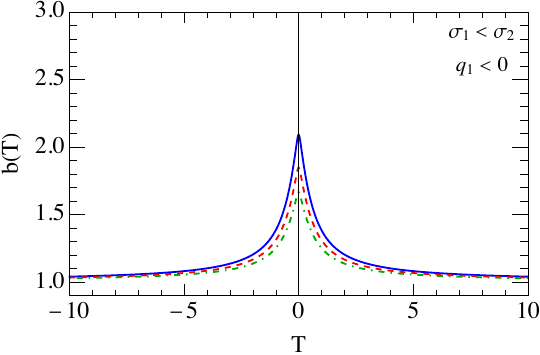}}

\caption{\label{fig1}{The scale factors to ordinary space, $a(T)$, and extra-dimensional space $b(T)$. The diamond symbol ($\diamond$) in Figs. (a) and (b) labels the scale factor $b(T)$.}}
\end{figure}

It is straightforward to verify that the symmetric bounce in the extra dimensions yields an asymmetric bounce in ordinary dimensions or vice versa.
Furthermore, the bouncing time in the extra coordinates differs from that in the brane hypersurface confined dimensions.
Note that we could define an effective cosmological constant that appears only in the quantum regime $\Lambda_{n}=-2{q_{1}}\sqrt{\frac{\eta_{+}|\omega|(n-5)}{(n-2)6}}$. In this case, $T\approx 0$, we have an effective anti-de Sitter Universe with a bounce ($q_1>0$) or an effective de Sitter Universe with a bounce ($q_1<0$). Since the bounce is symmetric for extra-dimensional space, $T_b=0$ is the bouncing time only for the higher-dimensional scale factor. In this case, we get the minimum or maximum value for the bouncing $b\left(0\right)=\exp\left[\frac{\sigma_1^2-\sigma_2^2}{4}\left(\sqrt{\frac{\eta_{+}|\omega|}{(n-4)(n-2)}}q_1+\sqrt{\frac{6}{(n-5)(n-4)}}\frac{H_a(+\infty)}{2}\right)\right]$.

The scale factors are shown in Fig. \ref{fig1}, where we have considered $H_a(+\infty)=1.0$ and $\omega=1.0$. In turn, Figs. \ref{fig1a} and \ref{fig1b} assume the same value for $\sigma_1=2.0$ and $\sigma_2=0.5$, whereas $q_1=2.0$ and $q_1=-2.0$, respectively. On the other hand, Figs. \ref{fig1c} and \ref{fig1d} consider $\sigma_1=0.5$ and $\sigma_2=2.0$ while again admitting $q_1=2.0$ and $q_1=-2.0$. It is now important to investigate the influence of the parameters $\sigma_1$, $\sigma_2$, and $q_1$ on the found solutions. First, we note that the $a(T)$ does not change its bouncing profile under the change of its parameters. On its turn, at $a(0)=\exp\left[\frac{\sigma_1^2+\sigma_2^2}{4}\left(\sqrt{\frac{\eta_{+}|\omega|(n-5)}{(n-2)6}}q_1+\frac{H_a(+\infty)}{2}\right)\right]$, which means that for higher $\left(\sigma_1^2+\sigma_2^2\right)$–values, we have an initially bigger three-dimensional Universe for a given $n$. However, when $T\rightarrow +\infty$, the $q_1$–parameter leads to the asymmetry of the bounce.

In the case where  $\sigma_1 > \sigma_2$ $(\sigma_1 < \sigma_2)$ and $q_1 > 0$ $(q_1 < 0)$, extra dimensions are bigger at the bouncing phase. After this period, these dimensions would shrink – hidden from detection – and reach the estimated value for today $(\mathcal{O}(\textrm{TeV})^{-1})$. On the other hand, when $\sigma_1 > \sigma_2$ $(\sigma_1 < \sigma_2)$ and $q_1 < 0$ $(q_1 > 0)$, the scale factor $b(T)$ gets its minimum value for $T=0$, growing from there to a finite value. In both cases, the stabilization of the extra dimensions has been achieved through a dynamic mechanism of compactification. One may note that, during the evolution of the Universe, it is possible that for $n=6$, extra dimension length could match the size of ordinary dimensions at particular times $T=T_{e_1}$ or $T=T_{e_2}$, which may be determined numerically. It then grows more than the usual dimensions over a cosmological time interval $(T_{e_1}<T<T_{e_2})$. Thus, one can see that, for specific time values $T=T_{e_1}$ or $T=T_{e_2}$, the Universe is isotropic, i.e., all directions have the same size.

At this point, it is convenient to rewrite the Hubble's parameters by substituting the constants \eqref{d1} and \eqref{d2} into Eqs. \eqref{Ha} and \eqref{Hb}, so we get
\begin{align}
H_{a}(T) = & \sqrt{\frac{2|\omega|(n-5)}{3(n-2)}}\left[2 T \left(\frac{1}{\sqrt{\sigma_{1}^4+16
   T^2}}+\frac{1}{\sqrt{\sigma_{2}^4+16 T^2}}\right)
   \left(\sqrt{\frac{3(n-2)}{2(n-5)| \omega |}} H_a(+\infty) \right.\right.\notag\\
   &\Biggl.\biggl. +\sqrt{\eta_{+}}q_{1}\Biggr) -\sqrt{\eta_{+}}q_{1}\Biggr],\label{Hanewn}
\end{align}
\begin{align}
H_{b}(T)= & \frac{1}{4}\sqrt{\frac{|\omega|}{(n-2)(n-4)}} \left(\frac{T}{\sqrt{\sigma_{1}^4+16
   T^2}}-\frac{T}{\sqrt{\sigma_{2}^4+16 T^2}}\right)\left(\sqrt{\frac{3(n-2)}{2(n-5)| \omega |}} H_a(+\infty) \right.\notag\\
   &\Bigg. +\sqrt{\eta_{+}}q_{1} \Bigg).\label{Hbnewn}
\end{align}

\begin{figure}[tb]
\subfloat[\label{fig2a}]{\includegraphics[scale=0.8]{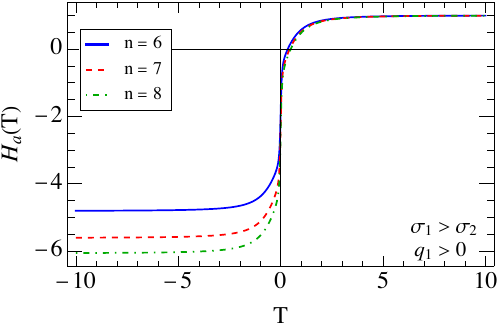}} \hspace{0.2cm} \subfloat[\label{fig2b}]{\includegraphics[scale=0.73]{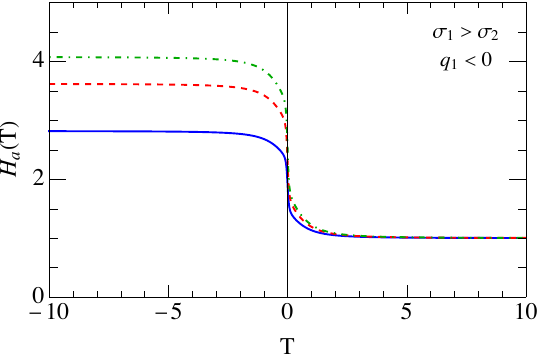}}

\subfloat[\label{fig2c}]{\includegraphics[scale=0.8]{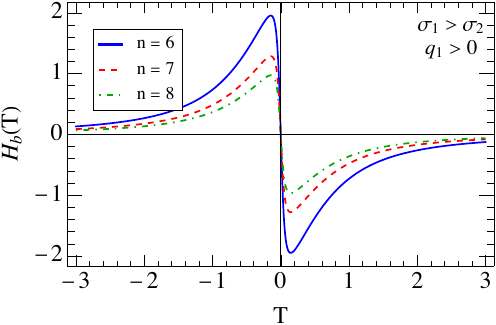}} \hspace{0.2cm} \subfloat[\label{fig2d}]{\includegraphics[scale=0.74]{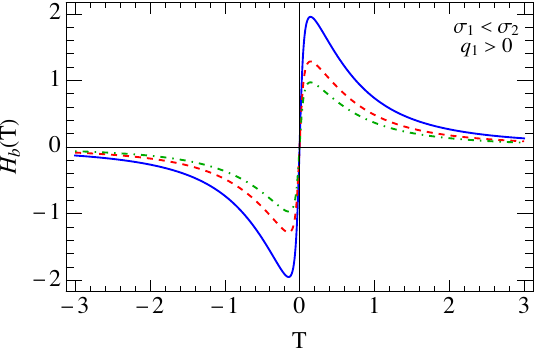}}

\subfloat[\label{fig2e}]{\includegraphics[scale=0.75]{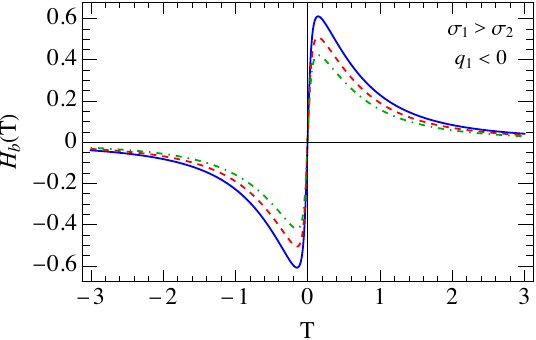}} \hspace{0.2cm} \subfloat[\label{fig2f}]{\includegraphics[scale=0.74]{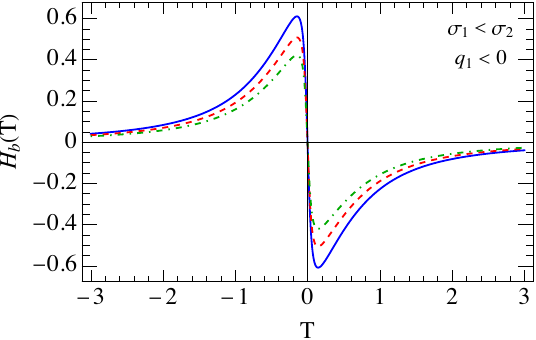}}
\caption{\label{fig2}{Hubble parameters as a function of cosmological time. We set $\sigma_1=2.0$, $\sigma_2=0.5$ with $q_1=2.0$ in Figs. \ref{fig2a} and \ref{fig2c}, whereas $q_1=-2.0$ in Figs. \ref{fig2b} and \ref{fig2e}. We assume $\sigma_1=0.5$ and $\sigma_2=2.0$ in Figs. \ref{fig2d} and \ref{fig2f}, where $q_1=2.0$, and  $q_1=-2.0$, respectively. Also, we keep set $H_a(+\infty)=1$ and $\omega=1$.}}
\end{figure}

The behavior of Hubble functions is shown in Fig. \ref{fig2}, where one assumes the same values for the constants such as those used in Fig. \ref{fig1}. It is easy to see that, although $H_{b}(T)$ is sensitive to change from $\left(\sigma_{1}>\sigma_{2} \right)$ to $\left(\sigma_{1}<\sigma_{2}\right)$, $H_{a}(T)$ remains unchanged despite these changes.
From Eq. \eqref{Hbnewn}, one can investigate the special condition that leads to $H_b(T)=0$: 
\begin{equation}
    q_{1}=-\frac{H_a(+\infty)}{2}\sqrt{\frac{(n-2)6}{\eta_{+}| \omega |(n-5)}}.
\end{equation}
It is worth highlighting that for this specific value of $q_{1}$, we find $H_a(T)=H_a(+\infty)$, which means that the quantum Universe enters an eternal de Sitter phase and, therefore, there is no bouncing phase. From \eqref{Hanewn}, we infer that for $q_1\neq 0$, the bouncing occurs before $T=0$, since $q_1>0$, or after $T=0$, once $q_1<0$. Only for $q_1=0$, the bouncing occurs at the origin. Further, we see from Fig. \ref{fig2} that a higher codimension yields a lower compactification rate for the extra
dimensions, while a lower codimension leads to a higher compactification rate.

It is also noteworthy that, by taking the classical regime in \eqref{a-final-n-dim} and \eqref{b-final-n-dim}, i.e., doing $T\rightarrow\pm \infty$, then $a(T)=e^{H_a(+\infty)T}$ and $b(T)=b_0$. Besides, we can note that $R^{n}\rightarrow R^{4}$ and the Universe behaves effectively as a four-dimensional hypersurface from the gravitational standpoint. In these solutions, we could also highlight an extra-dimensional era, representing a period in which the Universe has large extra dimensions than the ordinary dimensions. Qualitatively, we can see that from the Figs. \ref{fig1}.

\subsection{Non-Travelling Gaussian Wave Function $\Psi_{0}$}
Aiming to analyze the static Universe, here we take $q_1=0$,
which corresponds to the case where one constructs the Universe wave function using the initial wave packet $\Psi_0=\frac{\sqrt{2}}{\sqrt{\pi \sigma_1\sigma_2}}\exp\left(-\frac{\bar{x}^2}{\sigma_1^2}-\frac{\bar{y}^2}{\sigma_2^2}\right)$. So, the corresponding wave function of the Universe for all times is given by:
\begin{equation}\label{eq:scho-2}
 \Psi=\frac{1}{p_1p_2}\frac{\sqrt{2\sigma_1\sigma_2}}{\sqrt{\pi}}\exp\left(-\frac{\sigma_1^2\bar{x}^2}{16T^2+\sigma_1^4}-\frac{\sigma_2^2\bar{y}^2}{16T^2+\sigma_2^4}-\frac{4iT\bar{x}^2}{16T^2+\sigma_1^4}+\frac{4iT\bar{y}^2}{16T^2+\sigma_2^4}\right).  
\end{equation}
where $p_1=\sqrt{(i\sigma_1^2+4T)}$ and $p_2=\sqrt{-i\sigma_2^2+4T}$. Using the de Broglie-Bohm pilot waveform, $\Psi=Re^{iS}$, and rewritten $\bar{x}=\frac{x}{\sqrt{\eta_{+}}}$ and $\bar{y}=\frac{y}{\sqrt{|\eta_{-}|}}$ we can find the associated momenta by $p_x=\partial_xS$ and $p_y=\partial_yS$, which are equivalent to
\begin{equation}
    \dot{x}=\frac{16Tx}{16T^2+\sigma_1^4},
\end{equation}
\begin{equation}
\dot{y}=\frac{16Ty}{16T^2+\sigma_2^4}.
\end{equation}
Solving the equations and using the same boundary conditions employed previously, i.e., $H_b(T\rightarrow \infty)=0$ and $H_a(T\rightarrow \infty)=H_a(\infty)$ as a positive constant, we may arrive at the solutions \eqref{a-final-n-dim} and \eqref{b-final-n-dim} when $q_1=0$.
\begin{equation}
\label{a-final-n-dim-q=0}
a(T)=  \exp\left[\frac{H_a(+\infty)}{8}\left(\sqrt{\sigma_{1}^{4}+16T^{2}}+\sqrt{\sigma_{2}^{4}+16T^{2}}\right)\right],
\end{equation}
\begin{equation}
    \label{b-final-n-dim--q=0}
b(T)=\exp\left[\frac{H_a(+\infty)}{8}\sqrt{\frac{6}{(n-5)(n-4)}}\left(\sqrt{\sigma_{1}^{4}+16T^{2}}-\sqrt{\sigma_{2}^{4}+16T^{2}}\right)\right].
\end{equation}
One may note that, in terms of $A(T)=\frac{H_a}{2}\left(\sqrt{\frac{\sigma_1^4}{16}+T^2}+\sqrt{\frac{\sigma_2^4}{16}+T^2}\right)$, the bounce has, qualitatively, the same behavior of a 4-dimensional quantum Universe modeled with radiation $p=\rho/3$ \cite{delgado2020cosmological}, given that $a(T)$ is independent of the number of extra dimensions. In turn, in the case of the extra-dimensional scale factor, into $b(T)$ form, it seems to be a suitable way to look at the behavior of the Universe supplementary space. If $\sigma_1>\sigma_2$, there also exists a bounce, and the extra dimensions reach their maximum $b(0)$ at the bounce. Thus, for $T<0$, the extra dimensions have an increasing phase, while for $T>0$ have a decreasing phase.

On the other hand, for $\sigma_1<\sigma_2$, we have a typical bounce for which the extra dimensions reach their minimum at the bounce at $T=0$. For $T<0$, the extra dimensions are decreasing, while they are increasing for $T>0$. These two solutions are very similar to the cases indicated before. Nevertheless, there is no extra-dimensional era for $T>0$, which implies that the extra dimensions are always smaller than the usual spatial dimensions. The maximum or minimum is given by $b(0)=\exp\left[\frac{H_a(+\infty)}{8}\sqrt{\frac{6}{(n-5)(n-4)}}\left(\sigma_1^2-\sigma_2^2\right)\right]$. Furthermore, $b=1$ holds when $\sigma_1=\sigma_2$ or even as $T$ approaches infinity.

\section{Five-dimensional space-time}
From Eqs. \eqref{u} and \eqref{v}, we see that the case $n=5$ is problematic and requires separate treatment. For that, we start by considering the Sch\"{o}dinger-like equation for the five-dimensional Universe \eqref{psi_5}, and assume a new transformation $x=3A+B$ and $y=A-B$ so that we can arrive at the following Wheeler-DeWitt equation:
\begin{equation}
\left[ \frac{\partial^2 }{\partial x^2}-\frac{\partial^2}{\partial y^2}\right]\Psi=\frac{3i}{2\omega}\frac{\partial \Psi}{\partial T},
\end{equation}
This equation can be solved similarly to the previous sections, and one finds the scale factors under a change of variable $T\rightarrow \frac{2\omega}{3}\bar{T}$:
\begin{align}
a(\bar{T})=&\exp \Biggl\{\frac{1}{16}\left[\left(3H_{a}(+\infty)+2q_{1}\right)\sqrt{\sigma_{1}^4+16\bar{T}^2} +\left(H_{a}(+\infty)-2q_{2}\right)\sqrt{\sigma_2^4+16\bar{T}^2}\right.\notag\\
    &\left.\biggl.-8(q_1-q_2)\bar{T}\right]\Biggr\},
\end{align}
\begin{align}
    b(\bar{T})=&\exp \Biggl\{\frac{1}{16}\left[\left(3H_{a}(+\infty)+2q_{1}\right)\sqrt{\sigma_{1}^4+16\bar{T}^2} -3\left(H_{a}(+\infty)-2q_{2}\right)\sqrt{\sigma_2^4+16\bar{T}^2}\right.\notag\\
    &\left.\biggl.-8(q_1+3q_2)\bar{T}\right]\Biggr\},
\end{align}
where have been used the boundary conditions $\lim_{\bar{T}\to\infty}H_a(\bar{T})=H_a(+\infty)>0$, whereas $\lim_{\bar{T}\to\infty}H_b(\bar{T})=0$.

\begin{figure}[tb]
\subfloat[\label{fig3a}]{\includegraphics[scale=0.8]{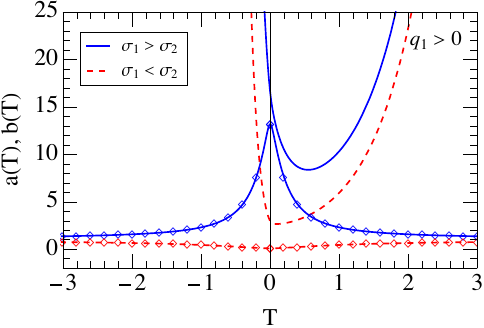}} \hspace{0.2cm} \subfloat[\label{fig3b}]{\includegraphics[scale=0.8]{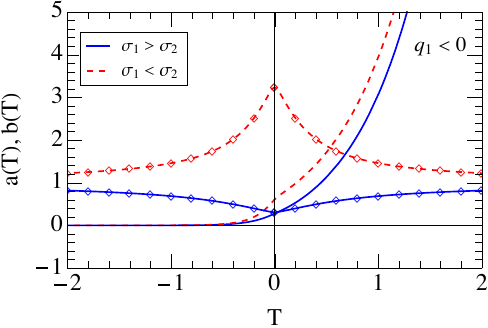}}

\caption{\label{fig3} In Fig. (a), we have been kept $q_1=4$ fixed, while in (b) $q_1=-4$. For both plots, $\sigma_i=2$ and $\sigma_j=0.5$, with $i,j=1$ or $2$. Again, diamond symbol ($\diamond$) in Figs. (a) and (b) identifies the scale factor $b(T)$.}
\end{figure}

It is worth highlighting that, as expected, $a(\bar{T})=e^{H_a\bar{T}}$ and $b(\bar{T})=b_0$ for the time $\bar{T}\rightarrow\infty$. Here we have to impose $q_2=-q_1/3$, to guarantee that the extra-dimensional cosmological bounce occurs at the origin. Under these assumptions, we can determine the Hubble parameters in a similar way to the previous section:
\begin{align}
H_{a}(\bar{T}) = & -\frac{2q_1}{3}+\frac{\left(3H_a(\infty) + 2q_{1}\right)}{3}\bar{T}\left(\frac{3}{\sqrt{\sigma_{1}^4+16
   \bar{T}^2}}+\frac{1}{\sqrt{\sigma_{2}^4+16 \bar{T}^2}}\right),\label{Hanew}
\end{align}
\begin{align}
H_{b}(\bar{T})= & \left(3H_a(\infty)+2q_{1}\right)\bar{T}\left(\frac{1}{\sqrt{\sigma_{1}^4+16
   \bar{T}^2}}-\frac{1}{\sqrt{\sigma_{2}^4+16 \bar{T}^2}}\right).\label{Hbnew}
\end{align}

\begin{figure}[tb]
\subfloat[\label{fig4a}]{\includegraphics[scale=0.8]{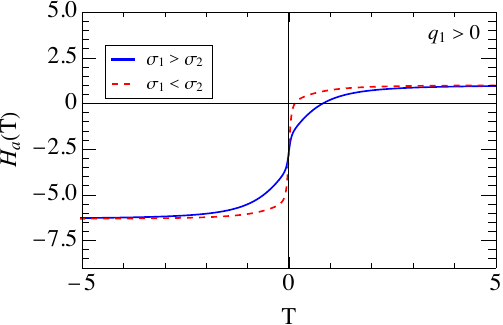}} \hspace{0.2cm} \subfloat[\label{fig4b}]{\includegraphics[scale=0.7]{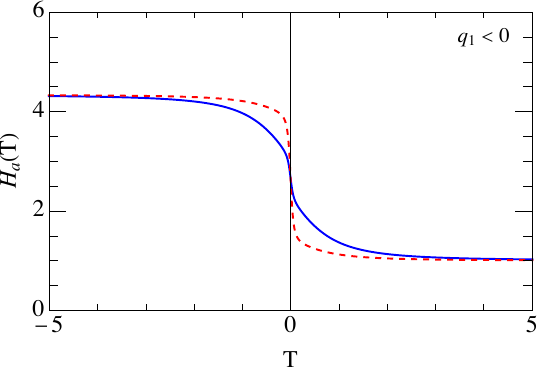}}

\subfloat[\label{fig4c}]{\includegraphics[scale=0.8]{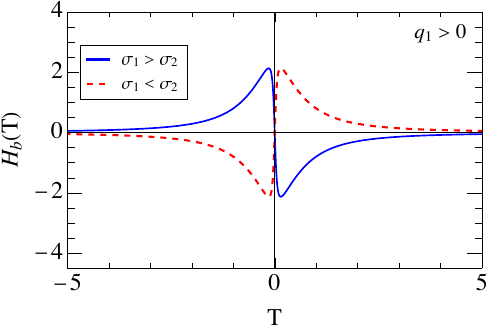}} \hspace{0.2cm} \subfloat[\label{fig4d.pdf}]{\includegraphics[scale=0.73]{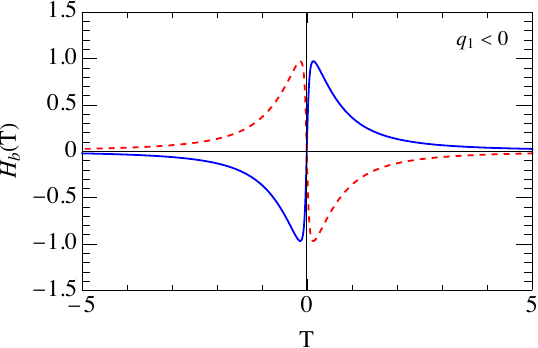}}
\caption{\label{fig4}{Hubble factors for $5$-dimensional Universe.}}
\end{figure}

At this point, we can determine a particular case that leads to $H_b(\bar{T})=0$: 
\begin{equation}
    q_{1}=-\frac{3}{2}H_a(+\infty),
\end{equation}
where $q_2=-q_1/3$. This particular solution describes an eternal static extra-dimensional Universe.
Notice that, for this specific $q_{1}$-value, we find $H_a(\bar{T})=H_a(+\infty)$. In turn, for $q_{1}<-\frac{3}{2}H_a(+\infty)$, non-physical solutions could be found since this condition leads to a shrinking four-dimensional Universe after the bounce.
Plots for scale factors and Hubble parameters, Figs. \ref{fig3} and \ref{fig4} were obtained by fixing $H_a(+\infty)=1$, $\omega=1$, $q_1=-4$, and returning to the original time-coordinate doing $\bar{T}\rightarrow \frac{3}{2\omega}T$. We set $q_1=4$ and $q_1=-4$ for Figs. (a), and (b), respectively.
From Fig. \ref{fig3}, we can see that for $q_1>0$ and $\sigma_1>\sigma_2$, the extra dimensions reach their maximum size at the bounce. Conversely, when $\sigma_2>\sigma_1$, they have their minimum size at the bounce. On the other hand, for the case $q_1<0$ and $\sigma_1<\sigma_2$, we found that the Universe could be so large that no period of inflation would be needed, along the same lines as \cite{pinto2009large}.
Further, if (i) $q_1<0$ and (ii) $H_a(+\infty)<\frac{2|q_1|}{3}$, so for $\bar{T} \leq	\frac{\sigma_2^2|3 H_a(+\infty)+2
   q_1|}{\sqrt{48 H_a(+\infty) (4 |q_1| - 3 H_a(+\infty))}}$, there will be $\bar{T}\ge0$ for which the extra dimensions will be larger than the ordinary dimensions, i.e., one gets an extra-dimensional era, such that $a(T)<b(T)$ in the entire quantum phase unless these two conditions, (i) and (ii), are not satisfied.

\section{Final Remarks}
In the present paper, we show extra-dimensional de Broglie-Bohm anisotropic Universes inspired by a generalization \cite{delgado2020cosmological}, where general bounce profiles have been studied. We show that bounces could occur for $k=0$, in contrast to what was found in the Randal-Sundrum-type model \cite{das2017bouncing}, in which extra-dimensional bounce only occurs if $k=-1$. 
A stabilization mechanism for the radii of the extra dimensions is required to ensure consistency between theoretical model predictions and empirical data for the fundamental physical constants (see for example \cite{Goldberger:1999uk,Bronnikov:2005iz} and references therein). Usually, the radion stabilization occurs when we define a strange matter \cite{carroll2002classical}, some noncommutative effect \cite{khosravi2007quantum,huang2001casimir}, or due to vacuum energy \cite{santos2008radion}.
It turns out that our solutions represent eternal Universe solutions in which a dynamical mechanism is responsible for stabilizing the extra-dimensional evolution.

Thus, we present several symmetrical extra-dimensional bounces investigated for $q_1$ and $q_2$ obeying a set relation. It is seen that all these solutions $b(T)$ show that stabilization occurs twice, before and after the bounce. Our solutions also reveal the possible existence of an extra-dimensional era, where the extra-dimensional Universe is larger than the $4D$ one. Although the $5$-dimensional Universe is governed by a different Wheeler-DeWitt equation, we show that the Universe follows what is expected from the general case ($n>5$).

The bounce solutions can be conveniently expressed in terms of cosmological quantities by relating the parameters of the wave function to observable quantities. For example, we can use the Kretschmann scalar $\mathcal{K}^{(n)}$ to define a scale for the bounce, $L_b$. For the sake of simplicity, for the case $q_1=0$ and $q_2=0$, we get $\mathcal{K}^{(n)}|_{T=0}= \frac{3 H^2_{a}(+\infty) b_{b}^{16-4 n}}{(n-5)a_{b}^{12} \sigma_{1}^4 \sigma_{2}^4 \omega^4}\left((n-3)\sigma_{1}^4+2 (n-7) \sigma_{1}^2\sigma_{2}^2+(n-3) \sigma_{2}^4\right)$. We can  define $L_b=1/\mathcal{K}^{1/4}|_{T=0}$. As a result, the scale of the quantum cosmology process taking the Planck scale demands that $L^{\left(n\right)}_{Pl}\leq L_b$. This constraint ensures the occurrence of bouncing solutions similar to what is found in Ref. \cite{delgado2020cosmological}, where the Universe is four-dimensional. Finally, we can highlight some future perspectives, such as studying the perturbations of our solutions aiming to investigate the stability of these Universes. Further, can the anti-matter or matter escape from the three-dimensional (observable) Universe in the extra-dimensional era? Is it necessary to impose some confinement mechanism to trap the ordinary matter? We intend to address these questions in future works.

\begin{acknowledgments}
We would like to thank CNPq, CAPES, and CNPq/PRONEX/FAPESQ-PB (Grant
No. $165/2018$) for partial financial support. F.A.B. acknowledges
support from CNPq (Grant No. 309092/2022-1). A.S.L. acknowledges support
from CAPES (Grant No. 88887.800922/2023-00).
\end{acknowledgments}


\providecommand{\noopsort}[1]{}\providecommand{\singleletter}[1]{#1}%
\begin{thebibliography}{10}

\bibliographystyle{ieeetr}

\bibitem{Nordstrom:1914ejq}
G.~Nordstrom, ``{On the possibility of unifying the electromagnetic and the
  gravitational fields},'' {\em Phys. Z.}, vol.~15, pp.~504--506, 1914.

\bibitem{Kaluza:1921tu}
T.~Kaluza, ``{Zum Unit\"atsproblem der Physik},'' {\em Sitzungsber. Preuss.
  Akad. Wiss. Berlin (Math. Phys. )}, vol.~1921, pp.~966--972, 1921.

\bibitem{Klein:1926tv}
O.~Klein, ``{Quantum Theory and Five-Dimensional Theory of Relativity. (In
  German and English)},'' {\em Z. Phys.}, vol.~37, pp.~895--906, 1926.

\bibitem{Randall:1999ee}
L.~Randall and R.~Sundrum, ``{A Large mass hierarchy from a small extra
  dimension},'' {\em Physical Review Letters}, vol.~83, pp.~3370--3373, 1999.

\bibitem{Randall:1999vf}
L.~Randall and R.~Sundrum, ``{An Alternative to compactification},'' {\em
  Physical Review Letters}, vol.~83, pp.~4690--4693, 1999.

\bibitem{Antoniadis:1998ig}
I.~Antoniadis, N.~Arkani-Hamed, S.~Dimopoulos, and G.~R. Dvali, ``{New
  dimensions at a millimeter to a Fermi and superstrings at a TeV},'' {\em
  Phys. Lett. B}, vol.~436, pp.~257--263, 1998.

\bibitem{Arkani-Hamed:1998jmv}
N.~Arkani-Hamed, S.~Dimopoulos, and G.~R. Dvali, ``{The Hierarchy problem and
  new dimensions at a millimeter},'' {\em Phys. Lett. B}, vol.~429,
  pp.~263--272, 1998.

\bibitem{Arkani-Hamed:1998sfv}
N.~Arkani-Hamed, S.~Dimopoulos, and G.~R. Dvali, ``{Phenomenology, astrophysics
  and cosmology of theories with submillimeter dimensions and TeV scale quantum
  gravity},'' {\em Physical Review D}, vol.~59, p.~086004, 1999.

\bibitem{Overduin:1997sri}
J.~M. Overduin and P.~S. Wesson, ``{Kaluza-Klein gravity},'' {\em Phys. Rept.},
  vol.~283, pp.~303--380, 1997.

\bibitem{wesson2006five}
P.~S. Wesson, {\em Five-dimensional physics: classical and quantum consequences
  of Kaluza-Klein cosmology}.
\newblock World Scientific, 2006.

\bibitem{chodos1980has}
A.~Chodos and S.~Detweiler, ``Where has the fifth dimension gone?,'' {\em
  Physical Review D}, vol.~21, no.~8, p.~2167, 1980.

\bibitem{candelas1984calculation}
P.~Candelas and S.~Weinberg, ``Calculation of gauge couplings and compact
  circumferences from self-consistent dimensional reduction,'' {\em Nuclear
  Physics B}, vol.~237, no.~3, pp.~397--441, 1984.

\bibitem{van1986spontaneous}
N.~Van~Hieu, ``Spontaneous compactification of extra dimensions in
  eleven-dimensional quantum gravity,'' {\em Fortschritte der Physik}, vol.~34,
  no.~7, pp.~441--455, 1986.

\bibitem{carroll2002classical}
S.~M. Carroll, J.~Geddes, M.~B. Hoffman, and R.~M. Wald, ``Classical
  stabilization of homogeneous extra dimensions,'' {\em Physical Review D},
  vol.~66, no.~2, p.~024036, 2002.

\bibitem{carroll2009dynamical}
S.~M. Carroll, M.~C. Johnson, and L.~Randall, ``Dynamical compactification from
  de sitter space,'' {\em Journal of High Energy Physics}, vol.~2009, no.~11,
  p.~094, 2009.

\bibitem{middleton2019anisotropic}
C.~Middleton, B.~A. Brouse~Jr, and S.~D. Jackson, ``Anisotropic evolution of
  d-dimensional frw spacetime,'' {\em The European Physical Journal C},
  vol.~79, no.~12, p.~982, 2019.

\bibitem{Novello:2009tn}
M.~Novello, J.~M. Salim, and F.~T. Falciano, ``{On a Geometrical Description of
  Quantum Mechanics},'' {\em Int. J. Geom. Meth. Mod. Phys.}, vol.~8,
  pp.~87--98, 2011.

\bibitem{pinto2021broglie}
N.~Pinto-Neto, ``The de broglie--bohm quantum theory and its application to
  quantum cosmology,'' {\em Universe}, vol.~7, no.~5, p.~134, 2021.

\bibitem{vicente2021quantum}
G.~Vicente, ``Quantum h{\v o}rava-lifshitz cosmology in the de broglie--bohm
  interpretation,'' {\em Physical Review D}, vol.~104, no.~10, p.~103525, 2021.

\bibitem{vicente2023bouncing}
G.~Vicente, R.~O. Ramos, and V.~N. Magalh{\~a}es, ``Bouncing and inflationary
  dynamics in quantum cosmology in the de broglie--bohm interpretation,'' {\em
  Physical Review D}, vol.~108, no.~2, p.~023517, 2023.

\bibitem{maniccia2024quantum}
G.~Maniccia and G.~Montani, ``Quantum gravity corrections to the inflationary
  spectrum in a bohmian approach,'' {\em Symmetry}, vol.~16, no.~7, p.~816,
  2024.

\bibitem{vakili2006compactification}
B.~Vakili, S.~Jalalzadeh, and H.~Sepangi, ``Compactification and signature
  transition in kaluza--klein spinor cosmology,'' {\em Annals of Physics},
  vol.~321, no.~11, pp.~2491--2503, 2006.

\bibitem{rodrigo1984wheeler}
E.~Rodrigo, ``The wheeler-dewitt equation and quantum kaluza-klein theories,''
  {\em Physics Letters A}, vol.~105, no.~4-5, pp.~196--198, 1984.

\bibitem{darabi2000quantum}
F.~Darabi, W.~Sajko, and P.~Wesson, ``Quantum cosmology of 5d non-compactified
  kaluza-klein theory,'' {\em Classical and Quantum Gravity}, vol.~17, no.~21,
  p.~4357, 2000.

\bibitem{jalalzadeh2003multi}
S.~Jalalzadeh, F.~Ahmadi, and H.~R. Sepangi, ``Multi-dimensional classical and
  quantum cosmology: exact solutions, signature transition and stabilization,''
  {\em Journal of High Energy Physics}, vol.~2003, no.~08, p.~012, 2003.

\bibitem{alves2018quantum}
F.~Alves-Junior, M.~Pucheu, A.~Barreto, and C.~Romero, ``Quantum cosmology in
  an anisotropic n-dimensional universe,'' {\em Physical Review D}, vol.~97,
  no.~4, p.~044007, 2018.

\bibitem{pandey2019anisotropic}
S.~Pandey, ``Anisotropic n-dimensional quantum cosmological model with fluid,''
  {\em The European Physical Journal C}, vol.~79, no.~6, p.~487, 2019.

\bibitem{das2017bouncing}
A.~Das, D.~Maity, T.~Paul, and S.~SenGupta, ``Bouncing cosmology from warped
  extra dimensional scenario,'' {\em The European Physical Journal C}, vol.~77,
  pp.~1--9, 2017.

\bibitem{anchordoqui2001brane}
L.~Anchordoqui, J.~Edelstein, C.~Nunez, S.~P. Bergliaffa, M.~Schvellinger,
  M.~Trobo, and F.~Zyserman, ``Brane worlds, string cosmology, and ads/cft,''
  {\em Physical Review D}, vol.~64, no.~8, p.~084027, 2001.

\bibitem{fabris1999quantum}
J.~C. Fabris, N.~Pinto-Neto, and A.~Velasco, ``Quantum cosmology in
  scalar-tensor theories with non-minimal coupling,'' {\em Classical and
  Quantum Gravity}, vol.~16, no.~12, p.~3807, 1999.

\bibitem{almeida2015quantum}
C.~R. Almeida, A.~B. Batista, J.~C. Fabris, and P.~R. Moniz, ``Quantum
  cosmology with scalar fields: self-adjointness and cosmological scenarios,''
  {\em Gravitation and Cosmology}, vol.~21, no.~3, pp.~191--199, 2015.

\bibitem{almeida2021quantum}
C.~R. Almeida, O.~Galkina, and J.~C. Fabris, ``Quantum and classical cosmology
  in the brans--dicke theory,'' {\em Universe}, vol.~7, no.~8, p.~286, 2021.

\bibitem{mukhanov2005physical}
V.~F. Mukhanov, {\em Physical foundations of cosmology}.
\newblock Cambridge university press, 2005.

\bibitem{vakili2012scalar}
B.~Vakili, ``Scalar field quantum cosmology: a schr{\"o}dinger picture,'' {\em
  Physics Letters B}, vol.~718, no.~1, pp.~34--42, 2012.

\bibitem{farajollahi2010dynamics}
H.~Farajollahi, M.~Farhoudi, and H.~Shojaie, ``On dynamics of brans--dicke
  theory of gravitation,'' {\em International Journal of Theoretical Physics},
  vol.~49, pp.~2558--2568, 2010.

\bibitem{schutz1971hamiltonian}
B.~F. Schutz~Jr, ``Hamiltonian theory of a relativistic perfect fluid,'' {\em
  Physical Review D}, vol.~4, no.~12, p.~3559, 1971.

\bibitem{bohm1952suggested}
D.~Bohm, ``A suggested interpretation of the quantum theory in terms of"
  hidden" variables. i,'' {\em Physical review}, vol.~85, no.~2, p.~166, 1952.

\bibitem{PhysRev.85.180}
D.~Bohm, ``A suggested interpretation of the quantum theory in terms of
  "hidden" variables. ii,'' {\em Phys. Rev.}, vol.~85, pp.~180--193, Jan 1952.

\bibitem{foo2022relativistic}
J.~Foo, E.~Asmodelle, A.~P. Lund, and T.~C. Ralph, ``Relativistic bohmian
  trajectories of photons via weak measurements,'' {\em Nature Communications},
  vol.~13, no.~1, p.~4002, 2022.

\bibitem{pinto2013quantum}
N.~Pinto-Neto and J.~Fabris, ``Quantum cosmology from the de broglie--bohm
  perspective,'' {\em Classical and Quantum Gravity}, vol.~30, no.~14,
  p.~143001, 2013.

\bibitem{anacleto2010acoustic}
M.~Anacleto, F.~Brito, and E.~Passos, ``Acoustic black holes from abelian higgs
  model with lorentz symmetry breaking,'' {\em Physics Letters B}, vol.~694,
  no.~2, pp.~149--157, 2010.

\bibitem{delgado2020cosmological}
P.~C.~M. Delgado and N.~Pinto-Neto, ``Cosmological models with asymmetric
  quantum bounces,'' {\em Classical and Quantum Gravity}, vol.~37, no.~12,
  p.~125002, 2020.

\bibitem{pinto2009large}
N.~Pinto-Neto, ``Large classical universes emerging from quantum cosmology,''
  {\em Physical Review D}, vol.~79, no.~8, p.~083514, 2009.

\bibitem{Goldberger:1999uk}
W.~D. Goldberger and M.~B. Wise, ``{Modulus stabilization with bulk fields},''
  {\em Physical Review Letters}, vol.~83, pp.~4922--4925, 1999.

\bibitem{Bronnikov:2005iz}
K.~A. Bronnikov and S.~G. Rubin, ``{Self-stabilization of extra dimensions},''
  {\em Physical Review D}, vol.~73, p.~124019, 2006.

\bibitem{khosravi2007quantum}
N.~Khosravi, S.~Jalalzadeh, and H.~Sepangi, ``Quantum noncommutative
  multidimensional cosmology,'' {\em General Relativity and Gravitation},
  vol.~39, pp.~899--911, 2007.

\bibitem{huang2001casimir}
W.-H. Huang, ``Casimir effect on the radius stabilization of the noncommutative
  torus,'' {\em Physics Letters B}, vol.~497, no.~3-4, pp.~317--322, 2001.

\bibitem{santos2008radion}
E.~Santos, A.~Perez-Lorenzana, and L.~O. Pimentel, ``{Radion stabilization from
  the vacuum on flat extra dimensions},'' {\em Physical Review D}, vol.~77,
  p.~025023, 2008.

\end{thebibliography}
\end{document}